\documentclass{ws-procs9x6}

\usepackage{graphicx}

\begin{document}

\title{SIVERS AND COLLINS EFFECTS IN\\ POLARIZED $pp$ SCATTERING PROCESSES\footnote{
Talk presented by F.~Murgia at the II International Workshop on Transverse Polarisation
Phenomena in Hard Processes (Transversity 2008), May 28-31, 2008, Ferrara (Italy).} }

\author{M.~ANSELMINO$^{\;\!1}$, M.~BOGLIONE$^{\;\!1}$, U.~D'ALESIO$^{\;\!2,3}$,\\
 E.~LEADER$^{\;\!4}$, S.~MELIS$^{\;\!1}$, F.~MURGIA$^{\;\!3,}$}

\address{1 -- Dipartimento di Fisica Teorica, Universit\`a di Torino and \\
          INFN, Sezione di Torino, Via P. Giuria 1, I-10125 Torino, Italy}
\address{2 -- Dipartimento di Fisica, Universit\`a di Cagliari,\\
Cittadella Universitaria di Monserrato, I-09042 Monserrato (CA), Italy}
\address{3 -- INFN, Sezione di Cagliari, C.P. 170, I-09042 Monserrato (CA), Italy}
\address{4 -- Imperial College London, Prince Consort Road, London SW7 2BW, U.K.}

\begin{abstract}
We summarize the present phenomenology of Sivers and Collins effects for transverse single spin asymmetries in polarized proton--proton collisions within the framework of the generalized parton model (GPM).
We will discuss a reassessment of the Collins effect and some preliminary predictions for SSA's in $p^{\uparrow} p\to\pi, K+X$ processes at RHIC obtained using updated information from SIDIS data and a new set of meson fragmentation functions.
\end{abstract}

\keywords{Single spin asymmetries, $pp$ collisions, Sivers and Collins effects}

\bodymatter

\section{Introduction}\label{intro}
Transverse single spin asymmetries (SSA) for inclusive single-particle production in polarized high-energy proton--proton collisions have historically motivated a huge theoretical and experimental effort to understand the role of spin and intrinsic parton motion in hadronic processes (see Ref.~[\refcite{D'Alesio:2007jt}] for an up-to-date review). At present, other asymmetries are experimentally at reach for meson production in semi-inclusive deeply inelastic scattering (SIDIS), in the Drell-Yan (DY) process, or in inclusive two-particle production in proton--proton collisions, which are considered safer from a theoretical point of view. Many leading-twist azimuthal asymmetries can be studied for such processes, in a context where factorization has been proven and some universality issues for the soft, transverse momentum dependent (TMD) functions involved, like the Sivers distribution and the Collins fragmentation function (FF), have been clarified.
The interplay among different theoretical models (the collinear twist-three formalism~[\refcite{Qiu:1998ia}] and TMD approaches) has also been analyzed, at least for some kinematical regimes.
SSA's in inclusive single-particle production in hadronic collisions look more involved. Being twist-three effects, several competing mechanisms may contribute. Factorization holds in the collinear twist-three approach~[\refcite{Qiu:1998ia}]; however, in phenomenological applications, leading order (LO) unpolarized cross sections are used, which often underestimate experimental data.
On the other hand, factorization has never been proven for TMD approaches, and has to be assumed.
Other open points concern the universality properties of the TMD functions.

Nevertheless, the study of SSA's in $p^\uparrow p\to h+X$ processes is still very important and useful. Together with the well-known fixed target results of the Fermilab-E704 collaboration, an increasing amount of challenging high-energy experimental data are becoming available from RHIC experiments, with enlarged statistics and kinematical coverage.
At this stage, a combined phenomenological analysis of these results with SIDIS and $e^+e^-$ data can give crucial information on the universality properties of the Sivers and Collins functions and on possible factorization-breaking terms in hadronic collisions in the kinematical regimes experimentally accessible.
\section{The generalized parton model}\label{gpm}
In our phenomenological analysis we adopt an LO pQCD approach including spin and intrinsic motion ($\boldsymbol{k}_\perp$) effects and all leading-twist polarized TMD distribution and fragmentation functions, which keep their simple partonic interpretation~[\refcite{Anselmino:2005sh}]. The complete $\boldsymbol{k}_\perp$ kinematics is included also in the partonic cross sections.
As already mentioned, for this approach factorization has not been proven yet, and has to be considered as a reasonable assumption deserving careful phenomenological tests. The same is true for the (assumed) universality of the TMD functions involved.

In the context of the generalized parton model and the helicity formalism the polarized differential cross section for the inclusive process $A(S_A)\, B(S_B)\to h + X$, where $A$, $B$, are two spin-1/2 hadrons ({\it e.g.} two protons) and $h$ an unpolarized particle ({\it e.g.} a meson or a photon), reads schematically (see Ref.~[\refcite{Anselmino:2005sh}] for details):
\begin{eqnarray}
&&d\sigma^{A(S_A)\,B(S_B)\to h+X} \propto \\
 &&\sum_{i=a,b,c} \sum_{\lambda^{}_i,\lambda'_i} \int dx_i\, d\boldsymbol{k}_{\perp i}\> \rho^{a/A}_{\lambda^{}_a,\lambda'_a}\,\hat{f}_{a/A,S_A}\, \rho^{b/B}_{\lambda^{}_b,\lambda'_b}\,\hat{f}_{b/B,S_B}\, M^{}_{\{\lambda^{}_i\}}M^*_{\{\lambda'_i\}}\, D^h_{\lambda^{}_c,\lambda'_c}\,.\nonumber
\label{master}
\end{eqnarray}
When summing over all partons and their helicities, all allowed combinations of TMD PDF's$\otimes$FF's$\otimes$polarized partonic cross sections appear~[\refcite{Anselmino:2005sh}].
By taking appropriate sums and differences of these cross sections (with $S_A$ or $S_B$ changing sign) all spin asymmetries can be written down. As an example, the transverse SSA, $A_N$, is given by $A_N = (d\sigma^\uparrow-d\sigma^\downarrow)/(d\sigma^\uparrow+d\sigma^\downarrow)$, where $d\sigma^{\uparrow,\downarrow}\equiv d\sigma^{A^{\uparrow,\downarrow}\,B\to h+X}$.
Notice that for the single-particle production case considered here, keeping at least the dominant $\boldsymbol{k}_\perp$ dependence in the partonic cross sections is crucial in order to get a non-vanishing SSA (this is not the case for double spin asymmetries). This also explains why, although involving leading-twist TMD PDF's and FF's, the global effect is higher-twist in $k_\perp/p_{hT}$. Again, this is not the case for inclusive double-particle production, where an observed low-energy scale is present ({\it e.g.} the unbalanced two-particle total transverse momentum).
\section{Collins effect in $pp$ collisions: a reassessment}\label{collins}
We recently realized that a sign mistake was present in our calculation of the SSA's in hadronic collisions in the $qg\to qg$ channel contribution to the Collins mechanism (see also Ref.~[\refcite{Yuan:2008tv}]).
This induced strong cancelations among the dominant partonic channels at large $x_F$, leading, together with non trivial effects due to azimuthal phase integrations, to a sizeable suppression of the Collins effect and a potential dominance of the Sivers contribution~[\refcite{Anselmino:2004ky}].
After correction this conclusion gets substantial change.
Although the role of azimuthal phases is confirmed, for E704 and RHIC kinematics the Collins effect is not strongly suppressed anymore. Therefore, it can play a substantial role and mix with the Sivers mechanism.

Let us however stress that the corrected sign mistake \emph{plays no role} in several phenomenological analyses recently performed by our group~(see Ref.~[\refcite{D'Alesio:2007jt}] and refs. therein). In particular, there is no change for:
1) The Sivers asymmetry in $p^\uparrow p\to h + X$ and in the Drell-Yan process, and for the $\Lambda$ transverse polarization in unpolarized hadronic collisions; 2) The azimuthal Sivers and Collins asymmetries in SIDIS processes (no gluonic channel is involved at LO); 3) The gluon Sivers effect in $p^\uparrow p\to D+X$ at RHIC energies. Furthermore, there are practically no changes for the bounds on the gluon Sivers function recently derived from the RHIC-PHENIX $p^\uparrow p\to\pi^0+X$ data at mid-rapidity.

\begin{figure}
\begin{center}
\includegraphics[width=1.6in,angle=-90]{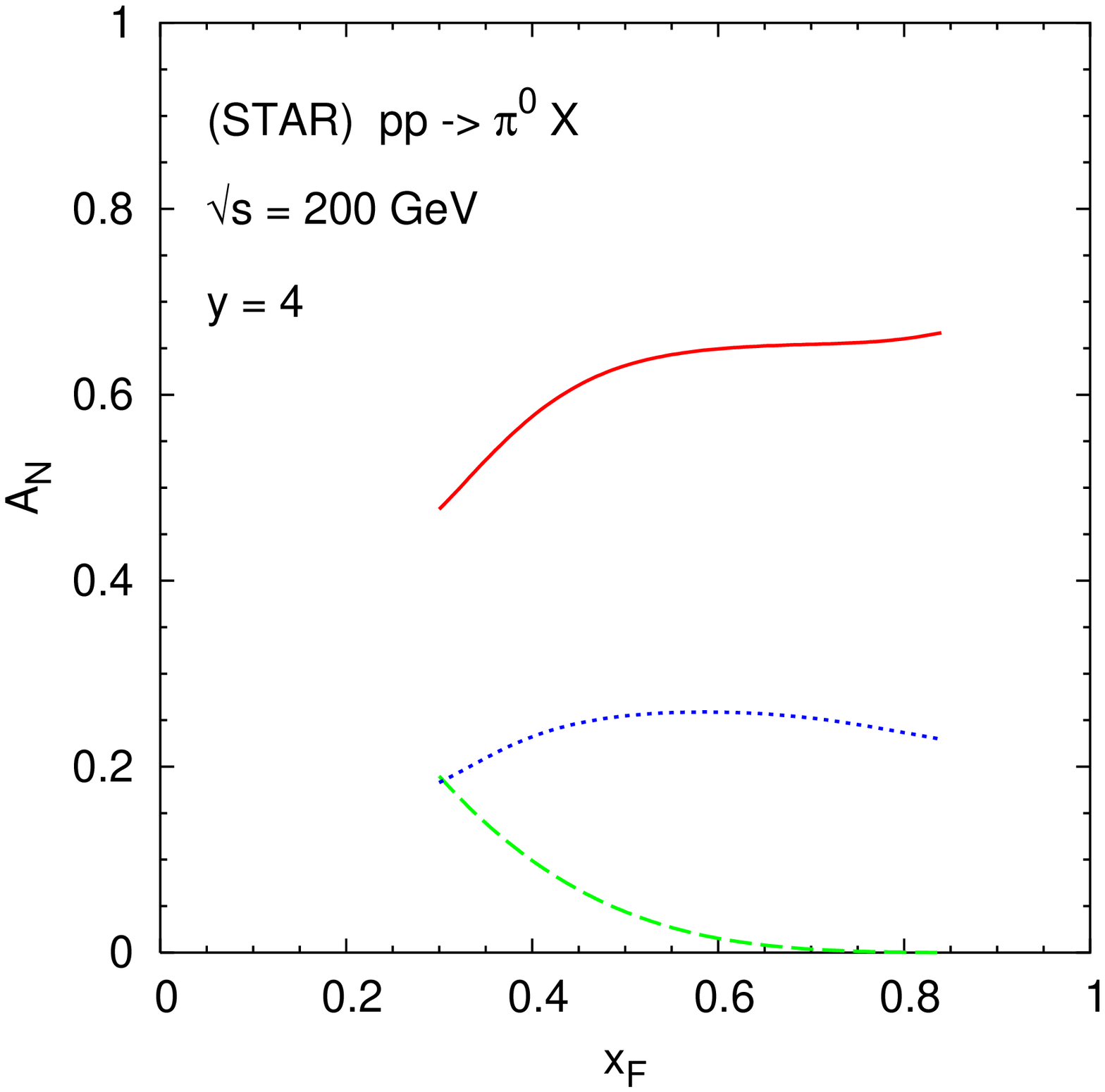}
\hspace*{20pt}
\includegraphics[width=1.6in,angle=-90]{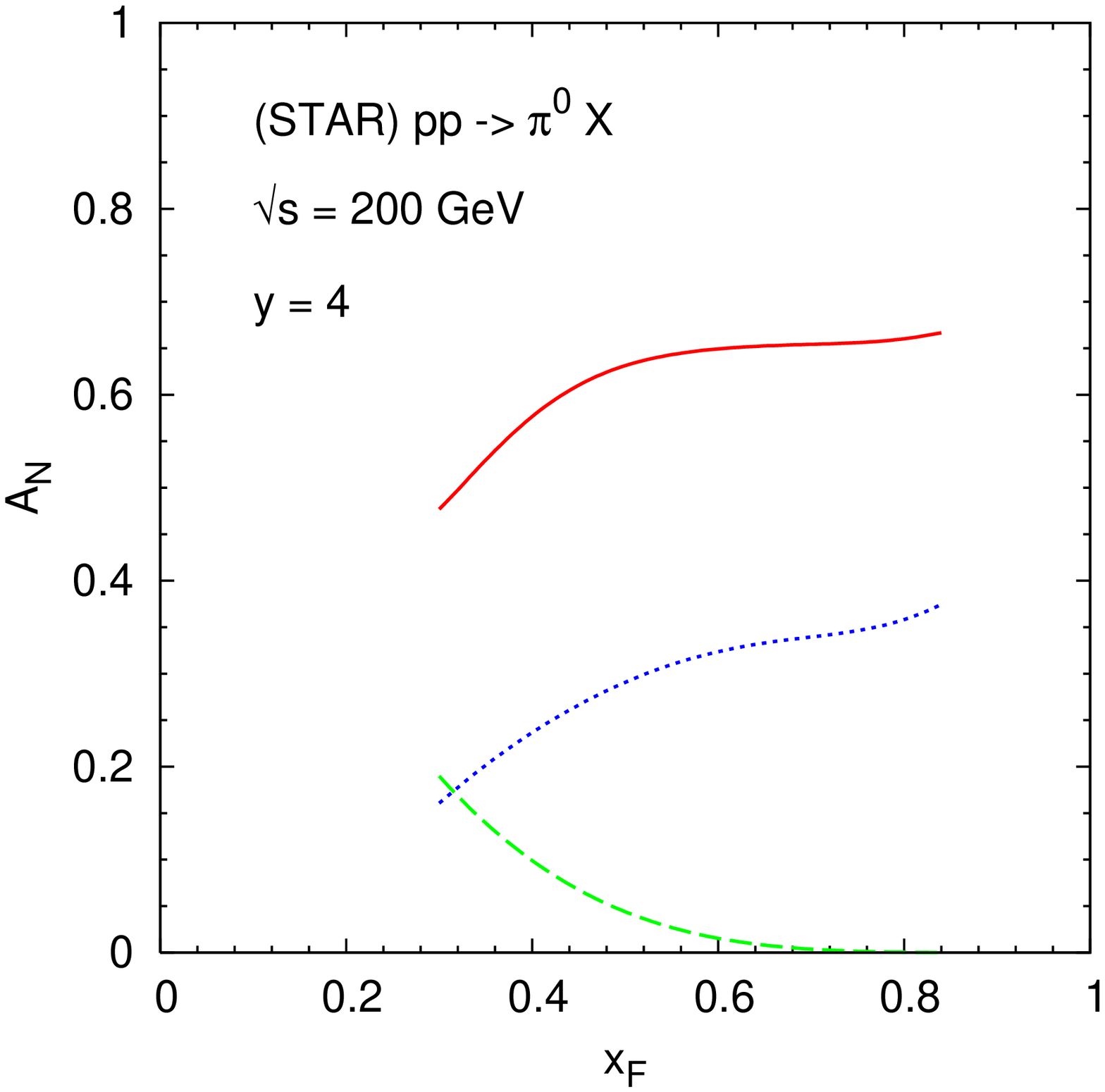}
\end{center}
\caption{Maximized contributions to $A_N(p^\uparrow p\to \pi^0+X)$ for STAR kinematics, $\sqrt{s}=200$ GeV, before (left) and after (right) correction: quark Sivers cont.~(solid line); gluon Sivers cont.~(dashed line); transversity$\otimes$Collins cont.~(dotted line). The PDF/FF sets are: MRST01/KKP.}
\label{coll-fig-star}
\end{figure}

\begin{figure}
\begin{center}
\includegraphics[width=1.2in,angle=-90]{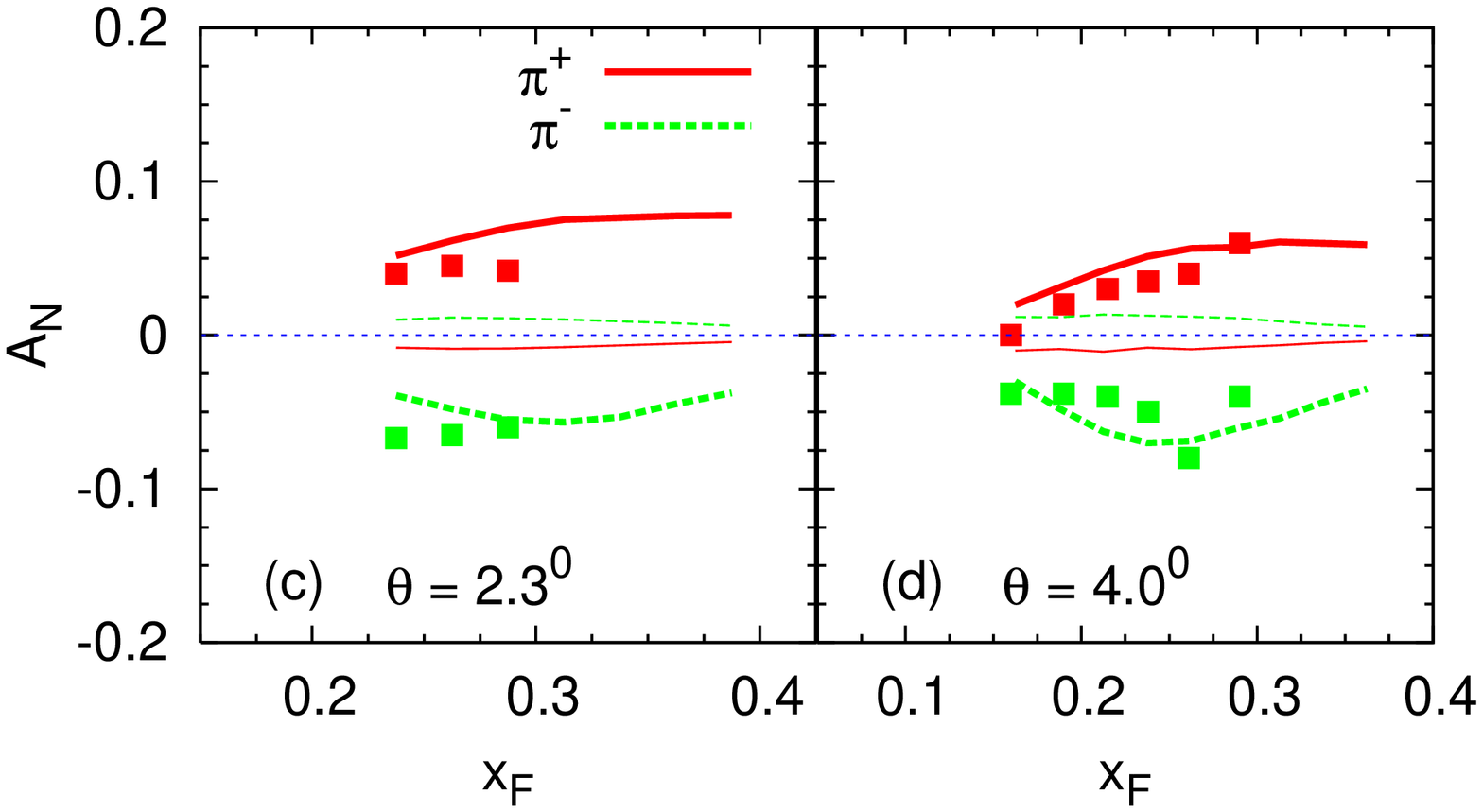}
\hspace*{-10pt}
\includegraphics[width=1.2in,angle=-90]{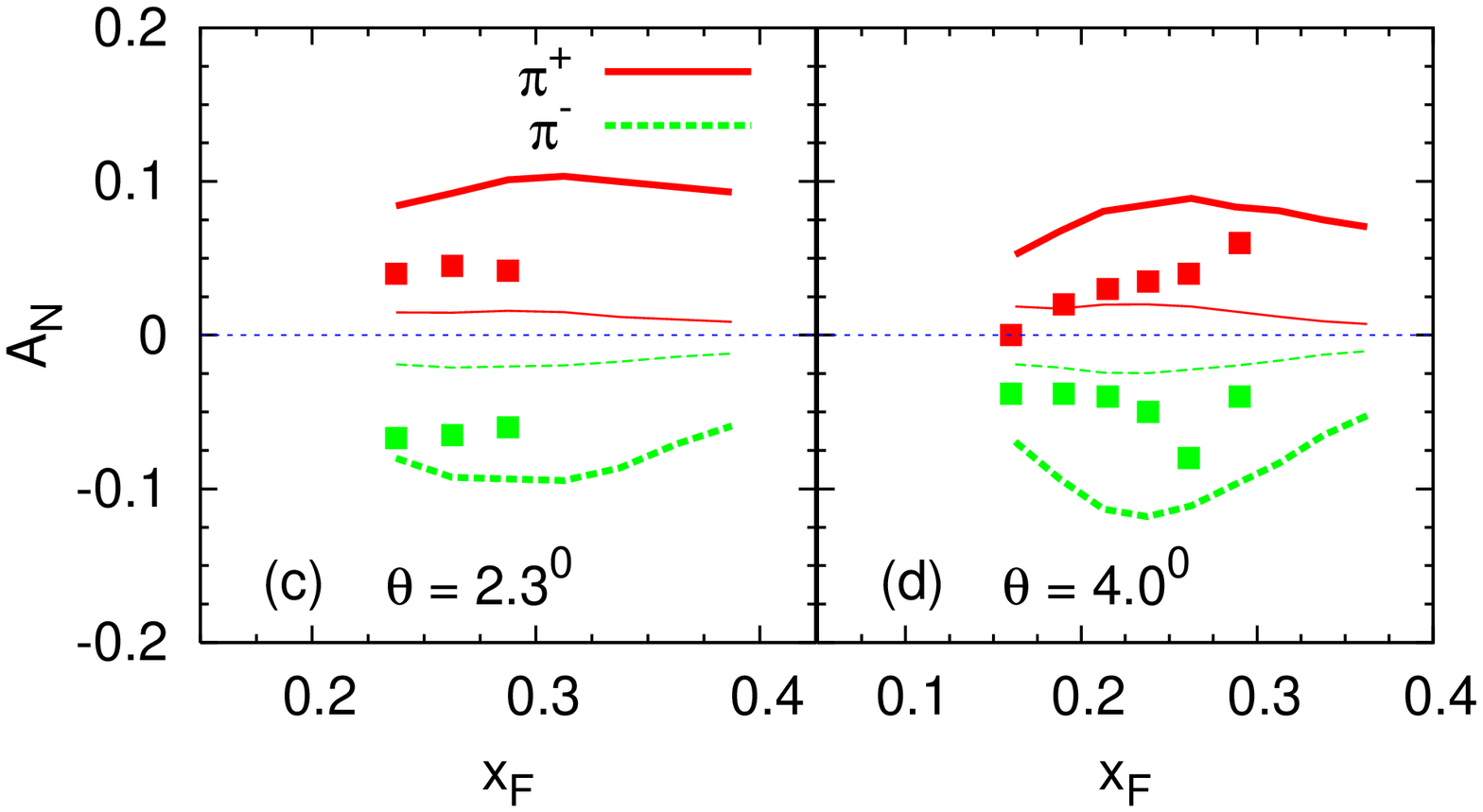}
\end{center}
\caption{Predictions for charged pion SSA's at RHIC-BRAHMS kinematics, $\sqrt{s}=200$ GeV, before (left) and after (right) correction. The parameterizations for the Sivers, transversity and Collins functions are from Refs.~[\refcite{Anselmino:2005ea},\refcite{Anselmino:2007fs}]. Thin lines: Collins contribution alone;
thick lines: Sivers + Collins contributions. The PDF/FF sets are: MRST01/Kretzer.}
\label{coll-fig-uni}
\end{figure}

In the rest of this section we will present a selection of the most significant variations in the phenomenological analyses which \emph{are} affected by this correction (an extended and comprehensive paper is in progress). The old results were obtained with parameterizations for the TMD PDF's and FF's which are not at present the most updated ones. In order to avoid confusion, and to clearly show the effect of the correction, we will present results using \emph{exactly} the same choices adopted in the original calculations. An analysis of the Collins effect in $pp$ collisions with the most up-to-date parameterizations is also in progress. Some preliminary results will be discussed in the next section.
In Ref.~[\refcite{Anselmino:2005sh}] we have considered the maximized contribution of the Collins mechanism to the transverse SSA for the process $p^\uparrow p\to\pi+X$ in the kinematical configurations of the E704 and STAR experiments.
\begin{figure}
\begin{center}
\includegraphics[width=1.6in,angle=-90]{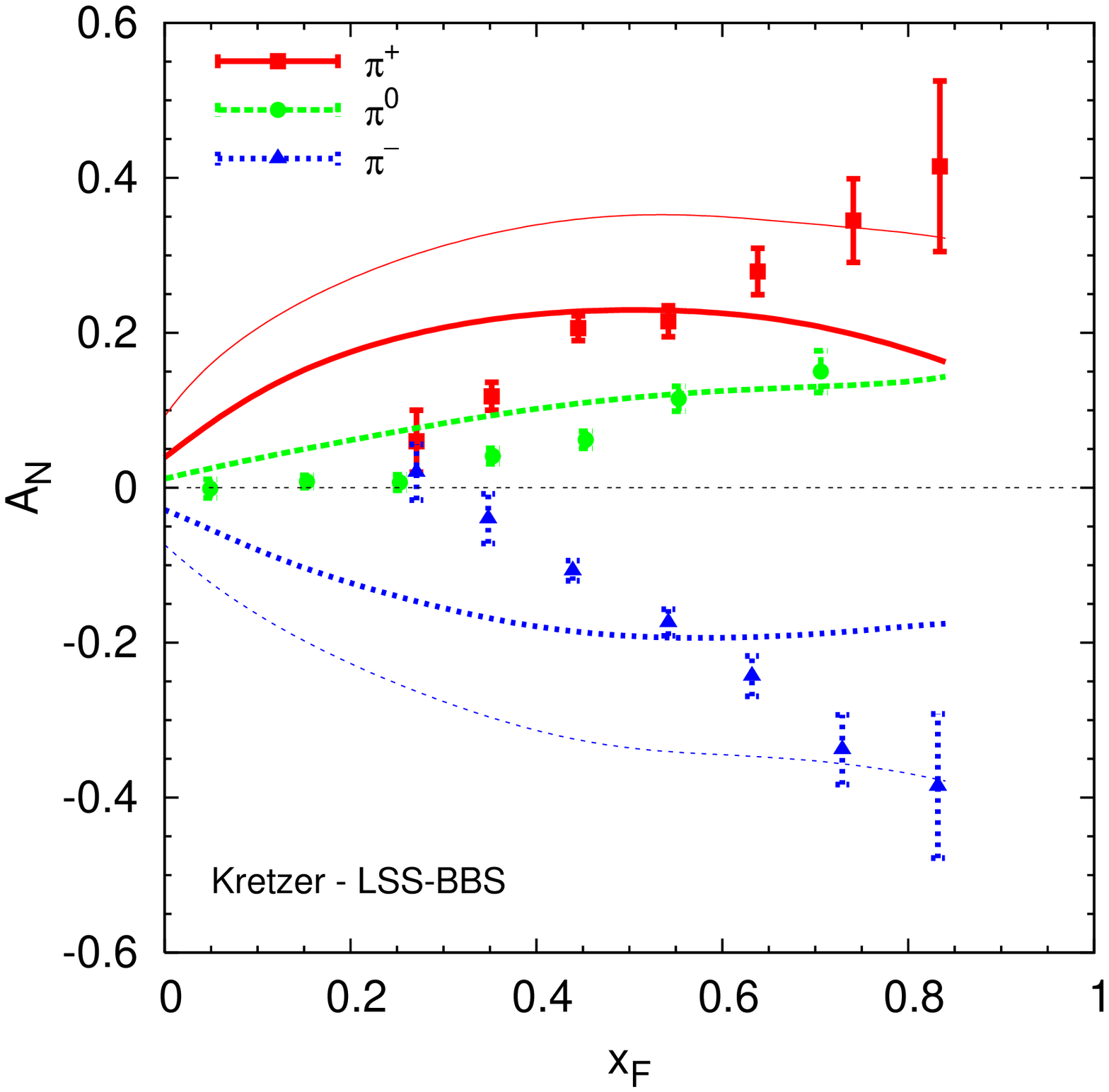}
\hspace*{20pt}
\includegraphics[width=1.6in,angle=-90]{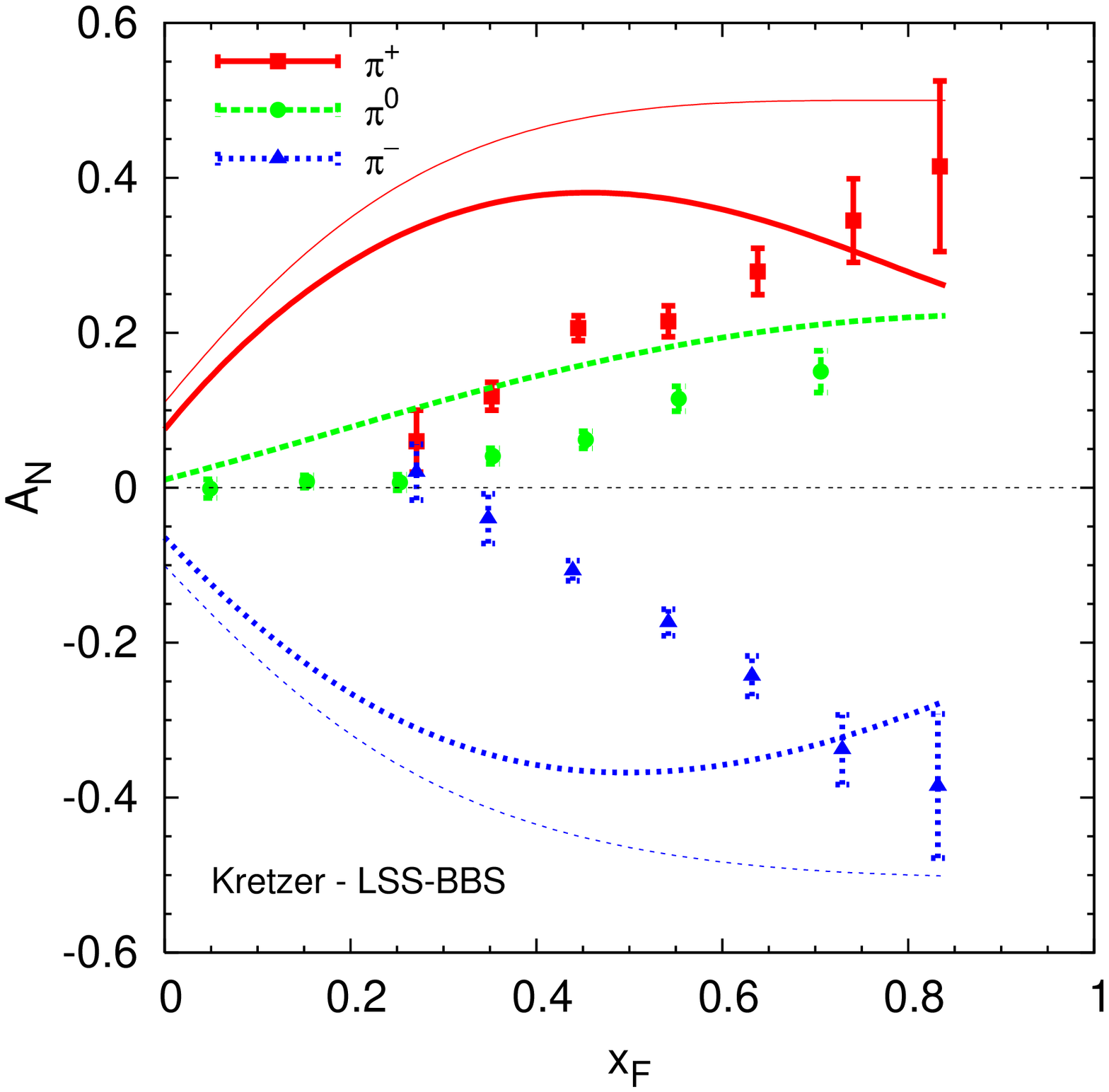}
\end{center}
\caption{Collins contribution (maximized effect) to pion SSA's for E704 kinematics, $\sqrt{s}=20$ GeV, before (left) and after (right) correction. Thin lines are obtained setting to zero all azimuthal phases. The PDF/FF sets are: LSS-BBS/Kretzer.}
\label{coll-fig-e704}
\end{figure}
In Fig.~\ref{coll-fig-e704} we compare the results for the E704 kinematics obtained before and after correction of the Collins contribution. Although the significant role of azimuthal phases in depleting the contribution at medium-large $x_F$ is confirmed, the suppression of the Collins effect at large $x_F$ is significantly reduced. Notice, however, that the plots show the maximized potential effect, with all partonic contributions saturated and added with the same sign.
Similarly, in Fig.~\ref{coll-fig-star} we compare results, before (Ref.~[\refcite{Anselmino:2005sh}]) and after correction, for the Sivers and Collins contributions to the $\pi^0$ SSA at the RHIC-STAR kinematics. Different maximized contributions are shown separately (quark Sivers, gluon Sivers and transversity$\otimes$Collins terms). Again, after correction the suppression of the maximized potential Collins contribution to the SSA is sizeably reduced at large $x_F$ (Sivers terms are clearly unaffected).
In a recent paper~[\refcite{Boglione:2007dm}], some of us have shown how predictions in very good agreement with RHIC data for pion SSA's can be obtained by using the central values of the parameterizations for the Sivers~[\refcite{Anselmino:2005ea}] and transversity distributions and the Collins function~[\refcite{Anselmino:2007fs}] then available from fits to the SIDIS and $e^+e^-$ data.
While the agreement with STAR data for $p^\uparrow p\to \pi^0+X$ is still good (see left plot in Fig.~\ref{newfits-star}), Fig.~\ref{coll-fig-uni} shows how for BRAHMS measurements our corrected predictions overestimate the data, somehow weakening the conclusions reached in Ref.~[\refcite{Boglione:2007dm}].

In conclusion, there is indeed a significant reduction of the Collins mechanism due to azimuthal phase factors. After correction, however, this effect alone is not sufficient to rule out sizeable Collins contributions to meson SSA's in $pp$ collisions, at least for the E704 and RHIC kinematics. Therefore, and in contrast with Ref.~[\refcite{Anselmino:2004ky}], we cannot exclude a more complicated (but even more interesting) situation in which both Sivers and Collins effects may potentially give significant and comparable contributions to single-particle SSA's in hadronic collisions.
\section{Sivers and Collins effects in $p^\uparrow\!\! p\to h+X$: a preliminary update}\label{newfits}
The analysis of the previous section was performed exactly in the same conditions as in the original papers. In the meantime both the amount and the precision of experimental data have significantly increased.
Moreover, a new set of meson fragmentation functions, the DSS set, has become available~[\refcite{deFlorian:2007aj}] which was obtained by including in the fit unpolarized cross sections and multiplicities for meson production in $pp$ collisions and in SIDIS. This set has a much larger $s$ quark component for kaon FF's and a much larger LO gluon component than previous sets, for both pions and kaons.

\begin{figure}
\begin{center}
\includegraphics[width=1.2in,height=2.2in,angle=-90]{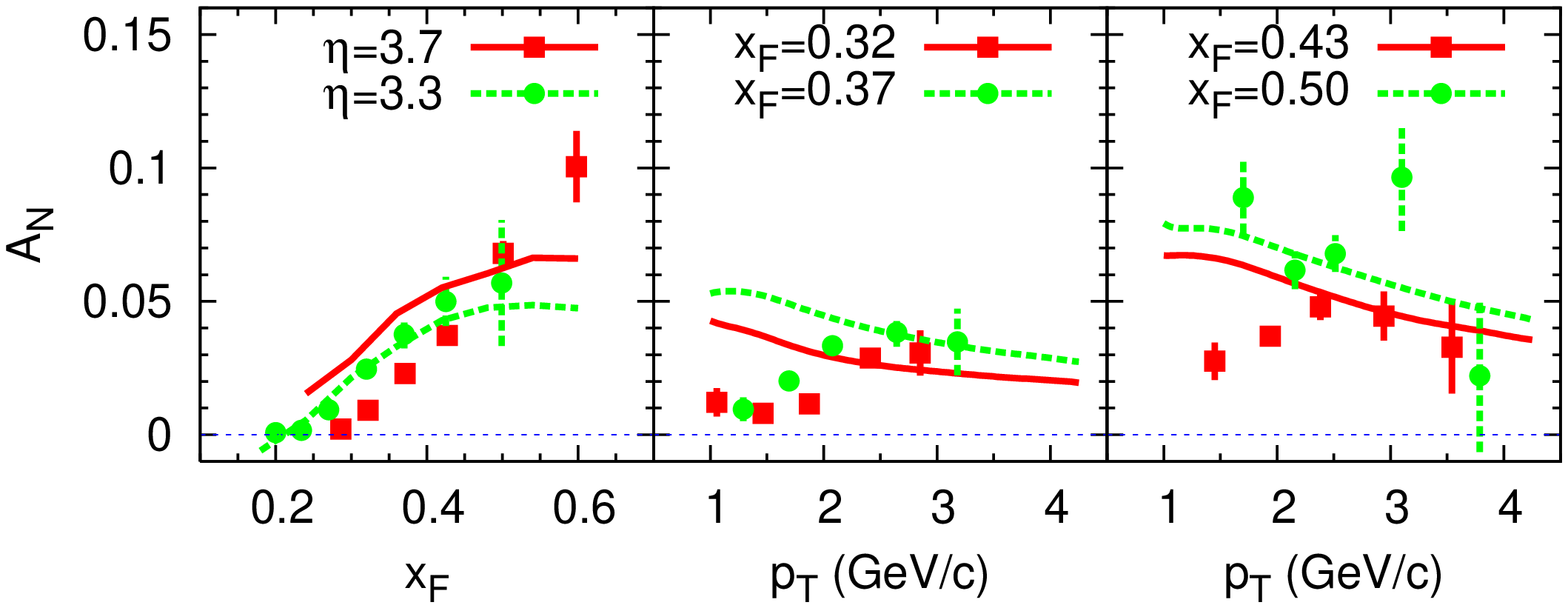}
\hspace*{-10pt}
\includegraphics[width=1.2in,height=2.2in,angle=-90]{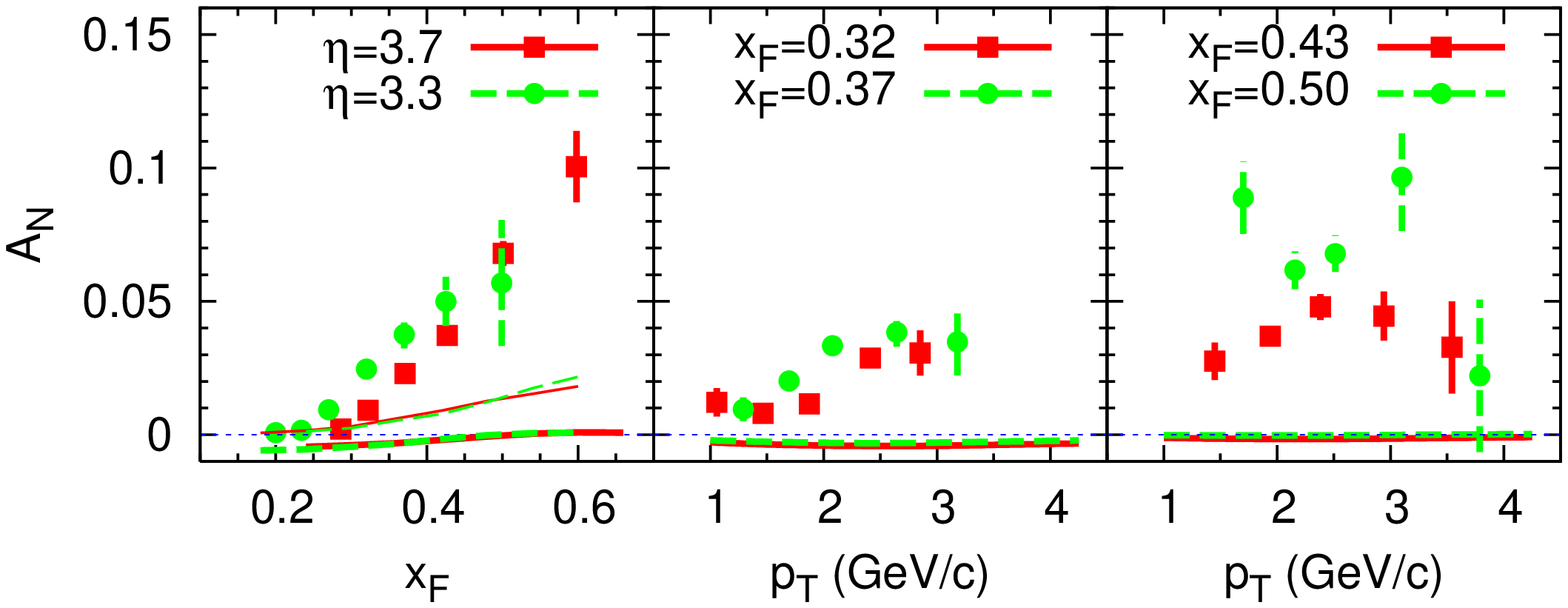}
\end{center}
\caption{Predictions for neutral pion SSA at RHIC-STAR kinematics, $\sqrt{s}=200$ GeV, obtained with the previous~[\refcite{Anselmino:2005ea},\refcite{Anselmino:2007fs}] (left) and new~[\refcite{Anselmino:2008sga},\refcite{Anselmino:2008sj}] (right) parameterizations for the Sivers, transversity and Collins functions. Left: Kretzer FF set, Collins + Sivers contributions;
right: DSS FF set, Sivers (thick lines) and Collins (thin lines, first panel only) contributions.
PDF set: MRST01.}
\label{newfits-star}
\end{figure}

\begin{figure}
\begin{center}
\includegraphics[width=1.2in,angle=-90]{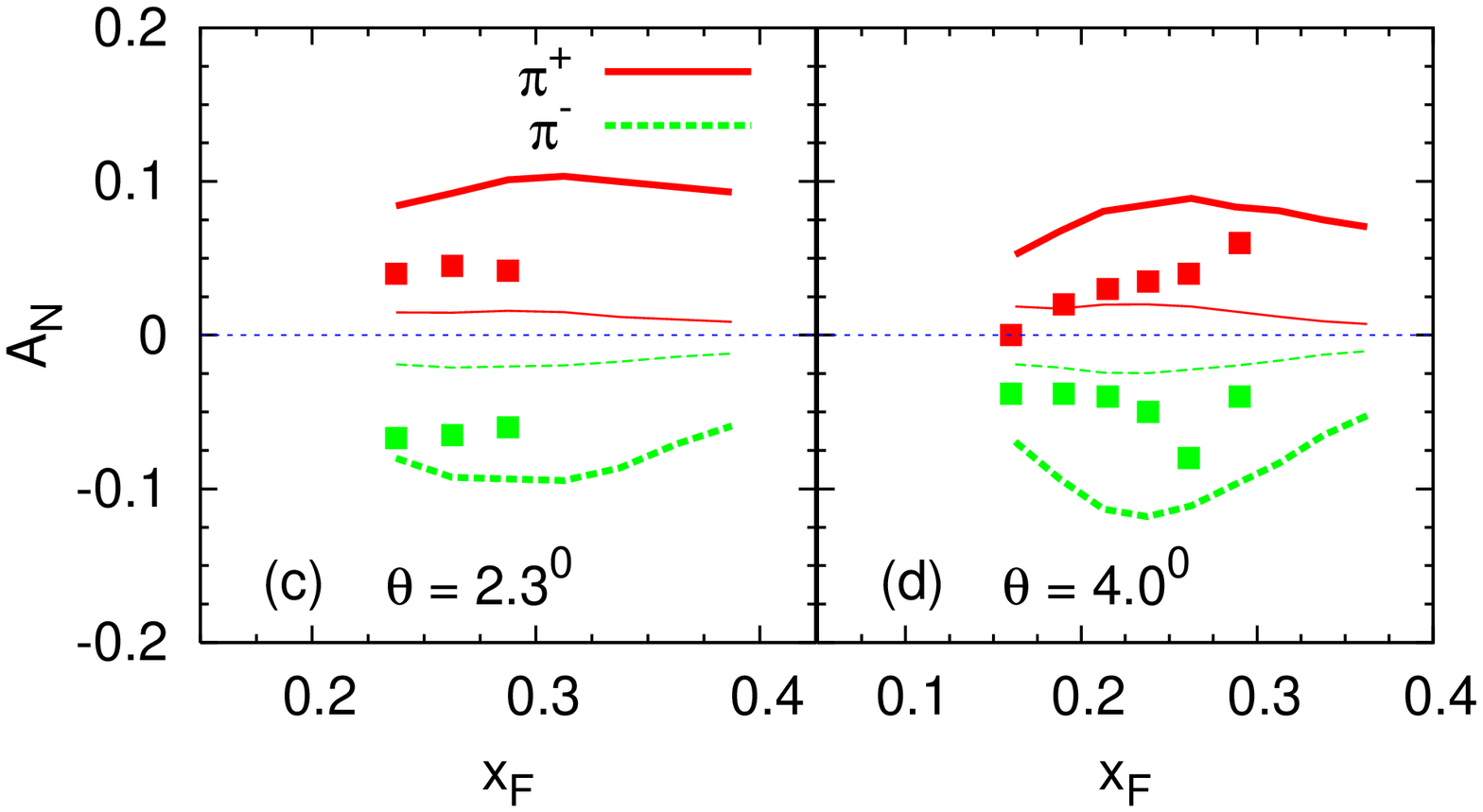}
\includegraphics[width=1.2in,angle=-90]{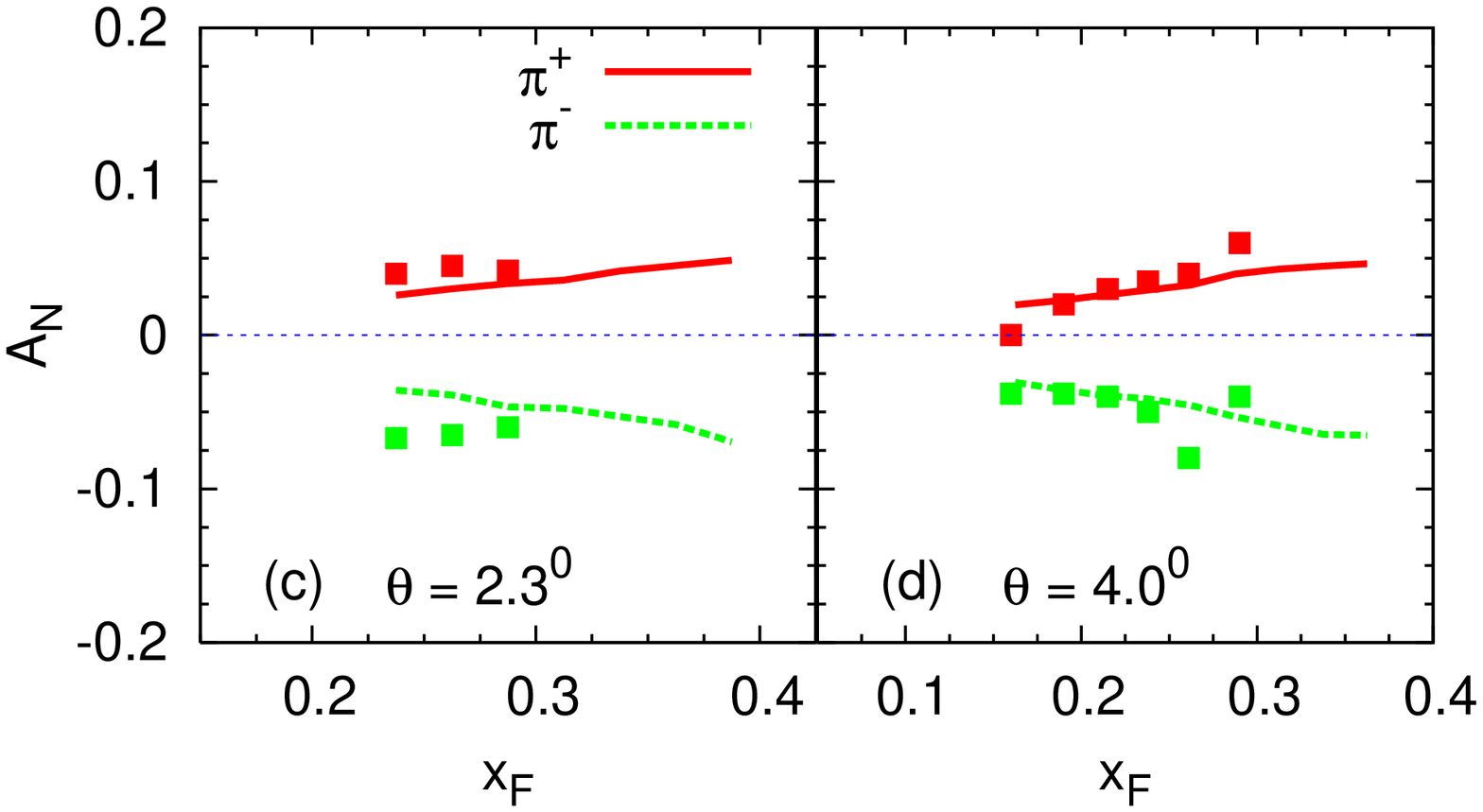}
\end{center}
\caption{Predictions for charged pion SSA's at RHIC-BRAHMS kinematics, $\sqrt{s}=200$ GeV, obtained with the previous~[\refcite{Anselmino:2005ea},\refcite{Anselmino:2007fs}] (left) and new~[\refcite{Anselmino:2008sga},\refcite{Anselmino:2008sj}] (right) parameterizations for the Sivers, transversity and Collins functions. Left: Kretzer FF set, Collins (thin lines) and Collins + Sivers (thick lines) contributions; right: DSS FF set, Collins + Sivers contributions. PDF set: MRST01.}
\label{newfits-brahms}
\end{figure}

The available SIDIS experimental data probe a (relatively) limited region of the Bjorken variable, $x_B$ ($x_B \leq 0.3-0.4$). The parameterizations of the TMD distributions involved are therefore almost unconstrained at large light-cone momentum fractions. This implies large uncertainties in the predictions for SSA's in $pp$ collisions at large positive $x_F$.
The new high-precision Belle data put instead more stringent constraints on the Collins FF.
The latest HERMES data on kaon SSA's are challenging, since positive kaons have on the average larger Sivers asymmetries than positive pions. This is difficult to understand in the simplified valence-like scenario adopted in the previous fits. Therefore, we have recently performed~[\refcite{Anselmino:2008sga}] a new fitting procedure for quark and, for the first time, sea-quark Sivers functions, using the DSS set and optimizing some parameters in order to: a) maximize the sensitivity in the $x_B$ region covered by the HERMES and COMPASS results; b) describe well the kaon data.
A new preliminary parameterization of the transversity and Collins functions, using the new SIDIS and $e^+e^-$ data is also available~[\refcite{Anselmino:2008sj}].
Details and features of the new fits and parameterizations have been already presented in this workshop (see the contributions of M.~Boglione and A.~Prokudin).
In Figs.~\ref{newfits-star} and \ref{newfits-brahms} we compare predictions for pion SSA's in $pp$ collisions at RHIC, using the corrected Collins contribution and the central-value parameterizations of the Sivers, Collins and transversity functions as obtained respectively in the previous (left plots) and in the new (right plots) fits. The results for STAR kinematics look much worse with the new fits, which however seem to reconcile (after correction of the Collins contribution) our predictions with the BRAHMS results (see Fig.~\ref{coll-fig-uni}).

These preliminary results must be taken with great care. As already discussed, SIDIS results do not constrain the large $x$ behaviour of the TMD distributions. As a consequence, predictions for SSA's at large $x_B$ or $x_F$ can be affected by huge uncertainties. They can also be strongly dependent on simplifying assumptions (necessary to reduce the number of free parameters in the fit) involving the parameters governing this large $x$ behaviour. In the new fits~[\refcite{Anselmino:2008sga,Anselmino:2008sj}], in the effort to probe the role of sea quarks in detail, we adopted the same large $x$ behaviour for all Sivers distributions. This is at variance with the old valence-like fits~[\refcite{Anselmino:2005ea,Anselmino:2007fs}], where the large $x$ behaviour of the $u$ and $d$ quark Sivers functions were left totally independent. The results of Fig.~\ref{newfits-star} clearly show that in order to have a sizable Sivers SSA at large $x_F$, the $u$ and $d$ quark Sivers functions \emph{cannot} have the same large $x$ behaviour. Notice that this is not in contradiction with the results of the new fits, since present SIDIS data are basically unaffected by these details.

Concerning the Collins contribution, it is indeed bigger after correction. However, for neutral pion production large cancelations between the leading and non-leading Collins functions (which result with opposite signs from the fits) prevent it to be very large. Let us also remark that the huge LO gluon component of the DSS set gives a large contribution to the denominator of the SSA but almost no contribution to its numerator (dominated by quark initiated terms).

This discussion allows us to understand the preliminary results of Figs.~\ref{newfits-star} and \ref{newfits-brahms}.
Clearly, in order to better investigate the possibility of simultaneously explain SIDIS and $pp$ data on meson SSA's with a universal set of TMD functions, we need to perform, in the lack of experimental SIDIS data at large $x_B$, a more detailed analysis. This is in progress and will be presented elsewhere.


\begin{thebibliography}{10}

\bibitem{D'Alesio:2007jt}
U.~D'Alesio and F.~Murgia, {\em Prog. Part. Nucl. Phys.} {\bf 61}, 394 (2008).

\bibitem{Qiu:1998ia}
J.-w. Qiu and G.~Sterman, {\em Phys. Rev. D} {\bf 59}, 014004 (1998).

\bibitem{Anselmino:2005sh}
M.~Anselmino {\em et~al.}, {\em Phys. Rev. D} {\bf 73}, 014020 (2006).

\bibitem{Yuan:2008tv}
F.~Yuan, {\em Phys. Lett. B} {\bf 666}, 44 (2008).

\bibitem{Anselmino:2004ky}
M.~Anselmino, M.~Boglione, U.~D'Alesio, E.~Leader and F.~Murgia,
 {\em Phys. Rev. D} {\bf 71}, 014002 (2005).

\bibitem{Boglione:2007dm}
M.~Boglione, U.~D'Alesio and F.~Murgia,
 {\em Phys. Rev. D} {\bf 77}, 051502(R) (2008).

\bibitem{Anselmino:2005ea}
M.~Anselmino {\em et~al.}, {\em Phys. Rev. D} {\bf 72}, 094007 (2005).

\bibitem{Anselmino:2007fs}
M.~Anselmino {\em et~al.}, {\em Phys. Rev. D} {\bf 75}, 054032 (2007).

\bibitem{deFlorian:2007aj}
D.~de~Florian, R.~Sassot and M.~Stratmann,
 {\em Phys. Rev. D} {\bf 75}, 114010 (2007).

\bibitem{Anselmino:2008sga}
M.~Anselmino {\em et~al.}, arXiv:0805.2677 [hep-ph] (2008).

\bibitem{Anselmino:2008sj}
M.~Anselmino {\em et~al.}, arXiv:0807.0173 [hep-ph] (2008).

\end{thebibliography}
\end{document}